\documentclass[12pt]{article}

\usepackage{amssymb}
\usepackage{mathtools}
\usepackage{amsmath}
\usepackage{amstext}
\usepackage{graphicx,epsfig}
\usepackage{epsfig}
\usepackage{verbatim} 	
\usepackage{caption}
\usepackage{fancybox}
\usepackage{slashed}
\usepackage{color}
\usepackage{ulem}
\usepackage{enumitem}
\usepackage{subfigure}
\usepackage{parskip}
\usepackage{dsfont}
\usepackage{tabu}
\usepackage{tikz}
\usepackage{booktabs}
\usepackage[numbers,sort&compress]{natbib}
\usepackage{cancel}

\usepackage{bbding}
\usepackage{wasysym}
\usepackage{multirow,hhline,array}
\usepackage[utf8]{inputenc}

\definecolor{navyblue}{rgb}{0.0, 0.0, 0.5}
\definecolor{royalblue}{rgb}{0.25, 0.41, 0.88}
\definecolor{cadmiumgreen}{rgb}{0.0, 0.42, 0.24}
\definecolor{blue-violet}{rgb}{0.54, 0.17, 0.89}
\definecolor{darkviolet}{rgb}{0.58, 0.0, 0.83}
\definecolor{orange(colorwheel)}{rgb}{1.0, 0.5, 0.0}

\usepackage{hyperref}
\hypersetup{colorlinks=true,linkcolor=royalblue,citecolor=magenta,pdfencoding=auto}
\usepackage{pifont}
\usepackage{dcolumn}
\usepackage{colortbl}
\newcommand{\Comment}[1]{{}}
\definecolor{MyDarkBlue}{rgb}{0.15,0.15,0.45}

\newcommand{\D}{{\rm d}}

\newcommand{\Mpl}{M_{\text{Pl}}}

\title{}
\author{}
{

\begin{document}

\begin{center}
{\Large\bf{Stable wormholes in scalar-tensor theories}}
\end{center} 

 \vspace{0.5truecm}
\thispagestyle{empty} 
\centerline{
Gabriele Franciolini${}^{a,}$\footnote{E-mail address: gabriele.franciolini@unige.ch},
Lam Hui${}^{b,}$\footnote{E-mail address: lh399@columbia.edu},
Riccardo Penco${}^{c,}$\footnote{E-mail address: rpenco@cmu.edu},
Luca Santoni${}^{b,}$\footnote{E-mail address: ls3598@columbia.edu},
Enrico Trincherini${}^{d,}$\footnote{E-mail address: enrico.trincherini@sns.it}
}

\vspace{0.5 cm}

\centerline{\it $^a$Department of Theoretical Physics and Center for Astroparticle Physics (CAP)}
\centerline{\it 24 quai E. Ansermet, CH-1211 Geneva 4, Switzerland}

\vspace{.1cm}

\centerline{\it $^b$Department of Physics, Center for Theoretical Physics, Columbia University,}
\centerline{\it New York, NY 10027, USA}

\vspace{.1cm}

\centerline{\it $^c$Department of Physics, Carnegie Mellon University, Pittsburgh, PA 15213, USA}

\vspace{.1cm}

\centerline{\it $^d$Scuola Normale Superiore, Piazza dei Cavalieri 7, 56126, Pisa, Italy and}
\centerline{\it INFN - Sezione di Pisa, 56100 Pisa, Italy}

\vspace{2cm}
\begin{abstract}

We reconsider the issue of whether scalar-tensor theories can admit stable wormhole configurations supported by a non-trivial radial profile for the scalar field.
Using a recently proposed effective theory for perturbations around static, spherically symmetric backgrounds, we show that scalar-tensor theories of ``beyond Horndeski'' type can have wormhole solutions that are free of ghost and gradient instabilities. Such solutions are instead forbidden within the more restrictive ``Horndeski'' class of theories.

\end{abstract}

\newpage

\section{Introduction}

Wormholes are interesting topological spacetime configurations that over the years have inspired not only the scientific literature. The original idea is usually traced back to the work by Einstein and Rosen \cite{PhysRev.48.73}, even if the interest in traversable wormholes as viable speculative shortcuts connecting different points in spacetime, rekindled only after the work by Morris and Thorne \cite{doi:10.1119/1.15620} (a chronological presentation of the early contributions can be found in \cite{Visser:1995cc}). Assuming a spherically symmetric spacetime geometry, they pointed out the need of some exotic form of matter with a non-standard equation of state to sustain the wormhole configuration and keep the wormhole throat open.
Indeed, it is well known that at classical level traversable wormholes are forbidden in General Relativity (GR) if the Null Energy Condition (NEC) is satisfied.  One way to overcome this obstruction is to consider systems where quantum effects compete with the classical ones in a controlled manner \cite{Visser:1995cc,Gao:2016bin,Maldacena:2018lmt,Caceres:2018ehr,Fu:2018oaq}, \textit{e.g.} in the case of charged massless fermions giving rise to a negative Casimir-like energy \cite{Maldacena:2018gjk}. Alternatively, one can consider additional fields with a non-trivial background profile which serves as a source of violation of the NEC already at classical level. Among the various microscopic candidates one could consider, arguably the simplest one is a single scalar field. 
Violations of the NEC are often associated with problematic instabilities~\cite{Dubovsky:2005xd,Rubakov:2014jja}. This is confirmed by several explicit examples \cite{ArmendarizPicon:2002km,Bronnikov:2002sf,Bronnikov:2002qx, Bronnikov:2006pt,Gonzalez:2008wd,Gonzalez:2008xk, Rubakov:2014jja, Rubakov:2015gza,Rubakov:2016zah, Evseev:2017jek} that reveal an instability---usually of the ghost-type---in the even sector for perturbations around the wormhole. These instabilities however are not unavoidable, as was shown in recent years in the context of FLRW cosmology~\cite{Creminelli:2016zwa,Cai:2016thi}.

Emboldened by these recent results, in this paper we will reconsider the stability of wormhole solutions in theories with a single scalar field coupled to gravity. More precisely, we will discuss the linear stability of such solutions both in the odd and in the even sector, focusing mostly on the positivity of the kinetic energy. We will also briefly comment on tachyonic instabilities in the odd sector.  To pursue such a program, we will use the Effective Field Theory (EFT) for static, spherically symmetric spacetimes recently introduced in \cite{Franciolini:2018uyq}. This framework has the advantage of being completely model-independent: it does not rely on any specific assumption regarding the explicit form of the covariant Lagrangian that defines the microscopic theory. This allows us to capture in general terms the origin of the ghost instability affecting the wormhole solutions found in the context of Horndeski theories \cite{Rubakov:2015gza,Rubakov:2016zah,Evseev:2017jek}\footnote{These no-go theorems are the counterparts of the ones originally derived in the context of FLRW cosmologies by~\cite{Libanov:2016kfc, Kobayashi:2016xpl}.} and show that such a pathology can be prevented by considering theories beyond the Horndeski class. Our work is a first step towards a more complete classification of the scalar-tensor theories that admit fully stable wormhole solutions. 

The rest of this paper is organized as follows. In Sec. \ref{sec:EFT} we briefly review the EFT for perturbations introduced in \cite{Franciolini:2018uyq}. We then discuss the background Einstein equations for a generic wormhole metric in Sec. \ref{sec:bckg}, and study the stability for both the even and odd sector in Sec. \ref{sec:eftw} before concluding in Sec. \ref{concls}. Additional technical details can be found in the appendices.

\noindent \textit{Conventions:} Throughout this paper we will work in units such that $c=\hbar=1$ and adopt a ``mostly plus'' metric signature. We will denote the scalar field with $\Phi$, to avoid any potential confusion with the angular variable $\phi$.

\noindent \textit{Note added:} While this paper was being finalized, we became aware of \cite{refId0}, which discusses an explicit example in the class of beyond-Horndeski theories that can support wormhole solutions that are stable,
a conclusion that is consistent with our findings.
However, our result indicates that stable wormhole solutions can be generically obtained without any fine-tuning, unlike in the explicit example discussed in~\cite{refId0}.
\\
After the submission of the first version of our paper, Ref.~\cite{Mironov:2018uou} appeared, fixing the error of~\cite{refId0} and the wrong conclusion about the fine tuning. Finally, the results of~\cite{Mironov:2018uou} fully agree with our findings.

\section{Effective theory for wormhole perturbations}
\label{sec:EFT}

In order to address the issue of stability of wormhole solutions in scalar-tensor theories, we will work within the EFT framework developed in~\cite{Franciolini:2018uyq}. There, we derived an effective action for perturbations around static, spherically symmetric backgrounds, which is valid provided the background configuration of the scalar field has a non-trivial radial profile. When that is the case, one can choose to work in ``unitary gauge'', where the scalar fluctuations vanish. The effective action can then be expressed solely in terms of metric fluctuations, and up to second order in derivatives and perturbations it reads\footnote{To be precise, the action \eqref{EFT-ss0} contains all the terms with at most two derivatives that contribute at quadratic order in perturbations. However, it also contains some terms of higher order in perturbations, which will not play any role in our discussion.}  
\begin{equation}
\begin{split}
& S =  \int\D^4x \, \sqrt{-g} \bigg[
\frac{1}{2}M^2_1(r) R -\Lambda(r) - f(r)g^{rr} - \alpha(r)\bar K_{\mu\nu} K^{\mu\nu}
\\
&	+ M_2^4(r)(\delta g^{rr})^2
	+M_3^3(r) \delta g^{rr }\delta K  + M_4^2(r) \bar K_{\mu\nu} \delta g^{rr }\delta K^{\mu\nu}
\\
&	+ M_5^2(r)(\partial_r\delta g^{rr})^2
	+M_6^2(r) (\partial_r\delta g^{rr})\delta K  + M_7(r)\bar  K_{\mu\nu} (\partial_r\delta g^{rr})\delta K^{\mu\nu}
	+ M_8^2(r)(\partial_a\delta g^{rr})^2
\\
&	+M_9^2(r)(\delta K)^2 + M_{10}^2(r)\delta K_{\mu\nu}\delta K^{\mu\nu} 
+ M_{11}(r) \bar K_{\mu\nu}\delta K \delta K^{\mu\nu} 
+ M_{12}(r) \bar  K_{\mu\nu}\delta K^{\mu\rho} \delta {K^{\nu}}_\rho 
\\
&	+ \lambda(r)\bar  K_{\mu\rho}{{\bar{K}}^{\rho}}_\nu \delta K \delta K^{\mu\nu}
+ M_{13}^2(r) \delta g^{rr } \delta\hat{R} 
+ M_{14}(r) \bar K_{\mu\nu} \delta g^{rr }\delta\hat{R}^{\mu\nu} + \ldots
\bigg] \, ,
\label{EFT-ss0}
\end{split}
\end{equation}
where $K_{\mu\nu} = \bar K_{\mu\nu} + \delta K_{\mu\nu}$ is the extrinsic curvature of surfaces of constant radius ($\bar K_{\mu\nu}$ denoting its background value, and $ \delta K_{\mu\nu}$ fluctuations around it).

The merit of the effective description \eqref{EFT-ss0} is that it is completely model independent, {\it i.e.} it does not make any assumption about the explicit form of the background solution or the covariant action describing the microscopic theory. The structure of the operators that enter the EFT \eqref{EFT-ss0} is dictated only by the spontaneous breaking of  $r$-translations due to the radial profile $\bar{\Phi}(r)$ of the scalar field. Any scalar-tensor theory in unitary gauge can be matched onto our effective action \eqref{EFT-ss0} with an appropriate choice of the EFT coefficients. In the absence of a specific model to match onto, though, these coefficient can be treated as arbitrary functions of the radial coordinate. 

Demanding that the background metric is a solution to Einstein equations, one can derive some constraints among the first few EFT coefficients in the effective action (see  below)~\cite{Franciolini:2018uyq}. These constraints are equivalent to the requirement that the action contains no tadpoles. In the absence of additional matter fields, one can also perform a conformal redefinition of the metric to set $M_1(r)\equiv\Mpl$. This simplifies considerably the effective action for perturbations as well as the constraints among the EFT coefficients.

\section{Wormhole background}
\label{sec:bckg}

We will now apply the general framework introduced above to the particular case of asymptotically flat Lorentzian wormholes. We will choose to work with a radial coordinate such that the background metric takes the form
\begin{equation}
\D s^2 = -a^2(r)\D t^2 + \D r^2 + c^2(r)\left(\D\theta^2+\sin^2\theta \, \D \phi^2 \right) \, ,
\label{GMN-2}
\end{equation} 
where the functions $a(r)$ and $c(r)$ obey the inequalities
\begin{equation}
a(r) \geq a_\text{min} >0 \, ,
\qquad
c(r) \geq c_\text{min} >0 \, ,
\label{backm1}
\end{equation}
and satisfy the asymptotic boundary conditions
\begin{equation}
a\rightarrow 1 \, ,
\qquad\quad 
c\rightarrow \pm r \, ,
\qquad\quad
\text{for } r\rightarrow \pm \infty \, .
\label{backm2}
\end{equation}
The quantity $c_\text{min}$ can be thought of as the radius of the wormhole throat.
These are the minimal requirements that a wormhole solution should satisfy. Additional criteria must be satisfied in order for such a solution to be truly traversable\footnote{For instance, requiring that tidal effects on the in-falling object are not too large may result in additional constraints on the wormhole metric functions and on the maximum speed of the traveller~\cite{Visser:1995cc}.}.

Notice that the condition \eqref{backm1} necessarily implies a violation of the null energy condition at the background level. This can be easily seen by considering the null vector $k_\mu = (a, 1,0,0)$ and calculating
\begin{equation}
 T_{\mu\nu} k^\mu k^\nu = \Mpl^2 G_{\mu\nu} k^\mu k^\nu = -2  \Mpl^2 \frac{a}{c}\left( \frac{c'}{a}  \right)' \, ,
 \label{necviola0}
\end{equation}
where $T_{\mu\nu}$ is the background energy momentum tensor associated with the last three tadpoles in the first line of Eq. \eqref{EFT-ss0}---see \cite{Franciolini:2018uyq} for the explicit expression.
From the result \eqref{necviola0}, one infers that the null energy condition $T_{\mu\nu} k^\mu k^\nu\geq0$ is violated at least in a neighborhood of the wormhole throat, where by definition $c'=0$ and $c''>0$.

Demanding that the metric \eqref{GMN-2} satisfies the background Einstein equations, $\Mpl^2 G_{\mu\nu}=T_{\mu\nu}$, amounts to imposing the following relations among the EFT coefficients $\alpha(r), f(r)$ and $\Lambda(r)$:
\begin{equation}
a\left[\left(\frac{a'}{a}-\frac{c'}{c} \right)\frac{\alpha}{a}\right]' 
+ \left( \frac{a''}{a} - \frac{c''}{c}  + \frac{a'c'}{ac}-\frac{c'^2}{c^2}+\frac{1}{c^2}  \right)\Mpl^2  =0 \, ,
\label{thirdtadpcondition}
\end{equation}
\begin{equation}
f(r) 
=  \left(\frac{a'c'}{ac} - \frac{c''}{c}  \right)\Mpl^2
- \left( \frac{3a'^2}{2a^2}-\frac{a'c'}{ac} +\frac{c'^2}{c^2} - \frac{a''}{2a}\right)\alpha + \frac{a'}{2a}\alpha' \, ,
\label{fm}
\end{equation}
\begin{equation}
\Lambda(r) 
= - \left(\frac{c''}{c} + \frac{a'c'}{ac}+\frac{c'^2}{c^2}-\frac{1}{c^2}  \right)\Mpl^2 
- \left( \frac{3a'^2}{2a^2}-\frac{a'c'}{ac} +\frac{c'^2}{c^2} - \frac{a''}{2a}\right)\alpha + \frac{a'}{2a}\alpha'
 \, .
\label{lm}
\end{equation}
Eq. \eqref{thirdtadpcondition} is a first order differential equation for $\alpha(r)$ with $r$-dependent coefficient, while Eqs. \eqref{fm}-\eqref{lm} fix $f(r)$ and $\Lambda(r)$ in terms of $\alpha(r)$ and the background metric. Notice that, in the particular class of theories where $\alpha\equiv0$, Eq. \eqref{thirdtadpcondition} reduces to a consistency relation for $a(r)$ and $c(r)$, which therefore are no longer independent:
\begin{equation}
 \frac{a''}{a} - \frac{c''}{c}  + \frac{a'c'}{ac}-\frac{c'^2}{c^2}+\frac{1}{c^2}  = 0 \, .
 \label{conseq0}
\end{equation}

\section{Stability of wormhole solutions}
\label{sec:eftw}

In this section, we assume that the underlying scalar-tensor theory admits a wormhole solution described by the background metric \eqref{GMN-2} satisfying the  conditions \eqref{backm1} and \eqref{backm2}, and we investigate the stability under perturbations. To this end, we expand the effective theory \eqref{EFT-ss0} up to quadratic order and study the dispersion relations of the physical fluctuations. The present discussion does not have the ambition to be a comprehensive classification of the operators that may or may not allow stable wormhole configurations. Instead, we are interested in showing that ({\it i}) the action \eqref{EFT-ss0} can be used to understand in general terms the origin of the pathology affecting many examples in the literature, and ({\it ii}) that it can be used to identify at least one of the operators among those listed in \eqref{EFT-ss0} that allows to evade the no-go restriction. As we shall see later on, the argument closely resembles the one for FLRW spacetimes \cite{Libanov:2016kfc, Kobayashi:2016xpl,Creminelli:2016zwa,Cai:2016thi}. Motivated by these results, we will consider a very small subset of the terms appearing in \eqref{EFT-ss0}, and  focus our attention on one particular two-derivative operator in the effective action \eqref{EFT-ss0}, namely $\delta g^{rr } \delta\hat{R}$. To be more precise, we will work with the following effective action:
\begin{equation}
 S^{(2)} =  \int\sqrt{-g} \left[
\frac{\Mpl^2}{2} R -\Lambda(r) - f(r)g^{rr}
	+ M_2^4(r)(\delta g^{rr})^2
	+M_3^3(r) \delta g^{rr }\delta K  
	+ M_{13}^2(r) \delta g^{rr } \delta\hat{R} 
\right] \, .
\label{bHorn-2}
\end{equation}

Two comments are in order at this point. First, we should stress that the choice \eqref{bHorn-2} it is not simply dictated by reasons of simplicity. In fact, some ``respectable'' covariant theories belonging to the class of quartic beyond-Horndeski theories \cite{Gleyzes:2014dya,Zumalacarregui:2013pma} reduce to \eqref{bHorn-2} in unitary gauge $\delta\Phi\equiv 0$, as shown explicitly in App. \ref{Sec:conf-trans}. Second, we emphasize that setting to zero all the other terms in \eqref{EFT-ss0} can be achieved without fine-tuning. It was in fact proven in \cite{Pirtskhalava:2015nla,Santoni:2018rrx} that a weakly broken galileon symmetry protects these couplings against large quantum corrections. In other words, any other operator in \eqref{EFT-ss0} is generated at quantum level at an energy scale that is parametrically larger than the one defining the tree-level interactions, allowing to set their couplings to zero in any practical application \cite{Pirtskhalava:2015nla,Santoni:2018rrx}. We can therefore safely assume that quartic beyond-Horndeski operators---see Eq. \eqref{0heb} in App. \ref{Sec:conf-trans}---are the only ones that significantly contribute at the second order in derivatives.

The generalization of our analysis to cases that include operators beyond the ones contained in \eqref{bHorn-2} is straightforward and beyond the scope of the present discussion.
Instead, in the following we will use the effective action \eqref{bHorn-2} to show that the wormhole background is always affected by a ghost instability if $M_{13}^2=0$ (in which case describes a theory in the Horndeski class \cite{Nicolis:2008in,Horndeski:1974wa, Deffayet:2011gz}). By contrast, we will see that when $M_{13}^2\neq0$ it is possible to avoid ghost and gradient instabilities alike.

Given the symmetries of the background, the metric perturbations can be decomposed into tensor harmonics~\cite{Regge:1957td} and divided into two sectors, based on whether they are even or odd under parity. These two sectors are decoupled at linear order and can be studied separately.\footnote{We are assuming here that parity is not broken neither explicitly nor spontaneously at the level of the fundamental theory.} Although the sector that is typically affected by ghost-like instabilities in Horndeski theories is the even one~\cite{Rubakov:2015gza,Rubakov:2016zah,Evseev:2017jek}, for the sake of completeness we will start by briefly reviewing the stability properties of the odd sector.

\subsection{Odd sector}

In this section, we will consider the odd sector, which in our case contains a single propagating degree of freedom. Following the procedure outlined in~\cite{Regge:1957td}, one can write the 
 most general odd-parity metric perturbation as:
\begin{equation}
\delta g_{\mu\nu} ^{\rm odd}= \begin{pmatrix}
0 &0  &- {\rm h}_0 \csc \theta  \partial_\phi & {\rm h}_0 \sin \theta  \partial_\theta  \\
0 & 0  & - {\rm h}_1 \csc \theta  \partial_\phi& {\rm h}_1 \sin \theta  \partial_\theta \\
- {\rm h}_0 \csc \theta  \partial_\phi & - {\rm h}_1 \csc \theta  \partial_\phi& \frac{1}{2} {\rm h}_2 \csc \theta \mathcal{X} &  -\frac{1}{2} {\rm h}_2 \sin \theta \mathcal{W}\\
 {\rm h}_0 \sin \theta  \partial_\theta&  {\rm h}_1 \sin \theta  \partial_\theta&  -\frac{1}{2} {\rm h}_2 \sin \theta \mathcal{W} &  -\frac{1}{2} {\rm h}_2 \sin \theta \mathcal{X} 
\end{pmatrix} Y_{\ell m} (\theta,\phi) e^{- i \omega t}\, , \label{odd metric perturbations}
\end{equation}
where ${\rm h}_0$, ${\rm h}_1$, ${\rm h}_2$ are functions of $r$, and where we have defined
\begin{equation}
\begin{aligned}
\mathcal{X} &= 2(\partial _ \theta \partial _ \phi - \cot \theta \partial _ \phi  )\,,
\\
\mathcal{W} &= (\partial _ \theta \partial _ \theta - \cot \theta \partial _ \theta  - \csc ^2 \theta  \partial _ \phi\partial _ \phi  )\,.
\end{aligned}
\end{equation}
Then, by performing an odd gauge transformation one can set ${\rm h}_2 = 0$. This is commonly referred to as the Regge-Wheeler gauge (see Sec. 5 of \cite{Franciolini:2018uyq} for further details).

As shown in \cite{Franciolini:2018uyq}, only the first three operators in the effective action~\eqref{bHorn-2} affect the odd sector, which is therefore the same as the one studied for instance in~\cite{ArmendarizPicon:2002km,Kobayashi:2012kh} by other methods. This sector is known to be stable, but will briefly review the main results within our formalism for completeness.

The EFT coefficients $f(r)$ and $\Lambda(r)$ are unambiguously determined by the Einstein equations in terms of $a(r)$ and $c(r)$---see Eqs. \eqref{fm} and \eqref{lm}. Therefore, the dynamics of the parity-odd sector is completely fixed once the background metric is specified. As discussed in \cite{ArmendarizPicon:2002km,Kobayashi:2012kh}, odd perturbations with $\ell =0$ do not exist, and with $\ell =1$ are pure gauge in the non-rotating case. The quadratic action for parity-odd modes with $\ell\geq2$ can be easily found by introducing an auxiliary field $\Psi$ and integrating both ${\rm h}_0$ and ${\rm h}_1$~\cite{Kobayashi:2012kh}.  The kinetic term in the odd sector of~\eqref{bHorn-2} then coincides with the one in GR, and therefore neither ghost nor gradient instabilities are present.

Additionally, one can derive an equation of motion for ${\rm h}_1$ in order to investigate the absence of tachyonic instabilities. After a change of coordinate of the form $\D\tilde{r}=\D r/a(r)$ and the field redefinition
\begin{equation}\label{odd-field-red}
 \Psi(\tilde r (r))\equiv \exp\left [  \int_{r_c} ^ r \left ( \frac{a'(l)}{a(l)}-\frac{c'(l)}{c(l)} \right )\D l  \right ]{\rm h}_1(r)\,,
\end{equation}
for some fiducial $r_c$, the equations of motion for the physical mode can be conveniently written in the form of a  Schr\"odinger-like equation \cite{Franciolini:2018uyq,ArmendarizPicon:2002km}:
\begin{equation}\label{sch-odd-e1}
	\left [ \frac{\D^2}{{\D \tilde r}^2}+ \omega^2  \right ]\Psi(\tilde r)=V(\tilde  r ) \Psi(\tilde r)\, 
\end{equation}
where the potential $V(\tilde r(r))$ reads
\begin{equation}\label{rwpot}
V(\tilde r(r))= - a^2 \left[ \frac{c'' (r)}{c (r)} - 2 \frac{c^{\prime 2} (r)}{c^2 (r)}+ 
	\frac{ a'(r) c'(r)}{a(r)c(r)}
	-\frac{(\ell+2)(\ell-1)}{c^2(r)}\right ]\, .
\end{equation}
After a quick inspection of the expression  \eqref{rwpot}, it is not hard to find explicit backgrounds such that the potential is manifestly positive everywhere, guaranteeing that the odd sector is also free of tachyonic instabilities at linear order. Indeed, let us consider for instance the case of a wormhole profile with $a\equiv1$ and $c^2(r)=r^2+c_\text{min}^2$ \cite{doi:10.1119/1.15620}, which trivially fulfils the condition \eqref{conseq0}. Then,  the potential \eqref{rwpot} takes on the form
\begin{equation}\label{rwpot2}
V(\tilde r(r))=  \ell(\ell+1) - 3 \frac{c_\text{min}^2}{(r^2+c_\text{min}^2)}  \, ,
\end{equation}
which is positive definite since $\ell\geq2$.


\subsection{Even sector}
\label{sec:l2}

Let us now turn our attention to the even sector. To this end, we parametrize the metric perturbations in the unitary gauge $\delta\Phi\equiv 0$ as follows \cite{Franciolini:2018uyq}:
\begin{equation}
\delta g_{\mu\nu}^\text{even} = \begin{pmatrix}
a^2H_0 & H_1  & \mathcal{H}_0 \partial_\theta & \mathcal{H}_0 \partial_\phi \\
H_1 & H_2  & \mathcal{H}_1 \partial_\theta & \mathcal{H}_1 \partial_\phi \\
\mathcal{H}_0 \partial_\theta & \mathcal{H}_1 \partial_\theta & c^2\left( \mathcal{K} + G \nabla_\theta \nabla_\theta \right) & c^2 G \nabla_\theta \nabla_\phi\\
\mathcal{H}_0 \partial_\phi & \mathcal{H}_1 \partial_\phi & c^2 G \nabla_\phi \nabla_\theta & c^2\left(\sin^2\theta \, \mathcal{K}+ G \nabla_\phi \nabla_\phi \right)  
\end{pmatrix} Y_{\ell m}(\theta,\phi) \, , \label{even metric perturbations}
\end{equation}
where $H_0$, $H_1$, $H_2$, $\mathcal{H}_1$, $\mathcal{H}_2$ $\mathcal{K}$ and $G$ are functions of $(t, r,\ell,m)$, and $\nabla_{\theta,\phi}$ are covariant derivatives on the $2$-sphere of radius one given by the following explicit expressions:
\begin{equation} \label{2 cov dev sphere}
\nabla_\theta \nabla_\theta = \partial_\theta^2 \, ,
\qquad
\nabla_\phi \nabla_\phi
= \partial_\phi^2 + \sin\theta \cos\theta \,  \partial_\theta 
\, ,
\qquad
\nabla_\theta \nabla_\phi = \nabla_\phi \nabla_\theta
= \partial_\theta \partial_\phi - {\cos\theta \over \sin\theta} \partial_\phi \, . 
\end{equation}
The spherical symmetry of the background guarantees that modes with different $\ell$, $m$ and parity are decoupled at linear order. Notice that the gauge fixing in the cases $\ell=1$ and $\ell=0$ deserve a separate discussion. For this reason, since the argument on the stability/instability of the wormhole solution does not depend on $\ell$, we decide to focus in the present section only on modes with $\ell\geq2$, collecting in App. \ref{App:l0l1}  all the expressions for $\ell=1$ and $\ell=0$ for the sake of completeness.

In the $\ell\geq2$ case, the remaining gauge freedom in \eqref{even metric perturbations} can be removed by fixing the residual temporal and angular diffeomorphisms in such a way that (see Sec. 6 of \cite{Franciolini:2018uyq})
\begin{equation}
\mathcal{H}_0 = G =0 \, ,
\end{equation}
which we will refer to as the Regge-Wheeler-unitary gauge. Thus, upon substituting into the action \eqref{bHorn-2} and expanding up to quadratic order in $\delta g_{\mu\nu}$, the Lagrangian for the physical modes can be found by following the steps outlined below (see also \cite{Kobayashi:2014wsa}).

After straightforward integrations by parts, it becomes manifest that the perturbation $H_0$ is a Lagrange multiplier. Equating its coefficient to zero yields the following condition:
\begin{multline}
H_2\left[
\frac{ c a' c'}{2a}   + \frac{c c''}{2}+ \frac{  c'^2}{2}+ \frac{\ell (\ell+1)}{4}+
\frac{M_3^3 \left( c^2 a'+2 a c c'\right)}{2 \Mpl^2 a}+
\frac{c^2 (M_3^3)'}{2 \Mpl^2}-\frac{\ell (\ell+1) M_{13}^2}{\Mpl^2}
\right]
\\
+H_2' \left( \frac{  c c' }{2} + \frac{  c^2 M_3^3}{2 \Mpl^2} \right)
+\frac{1}{4} \left(\ell^2+\ell-2\right) \mathcal{K}
-\frac{3}{2} c c' \mathcal{K}'
-\frac{1}{2} c^2 \mathcal{K}''
\\
- \mathcal{H}_1 \frac{\ell (\ell+1) c'}{2 c}
-\frac{1}{2} \ell (\ell+1)  \mathcal{H}'_1 =0 \, .
\label{LM1}
\end{multline}
Then, using the field redefinition
\begin{equation}
H_2 \equiv \left( \frac{  c c' }{2} + \frac{  c^2 M_3^3}{2 \Mpl^2} \right)^{-1}
\left[
\psi + \frac{1}{2} c^2 \mathcal{K}' + \frac{1}{2} \ell (\ell+1)  \mathcal{H}_1 
\right] \, ,
\label{psi}
\end{equation}
Eq. \eqref{LM1} becomes an algebraic equation for $\mathcal{H}_1$, which can be solved in terms of $\mathcal{K}$, $\psi$ and their derivatives.

Furthermore, one can integrate out $H_1$ upon using its equation of motion, which reads
\begin{equation}
\frac{1}{2} \ell(\ell+1) H_1
+\dot{H}_2\left(c c'+\frac{c^2 M_3^3}{\Mpl^2}\right)
+\dot{\mathcal{K}} \left(\frac{c^2 a'}{a}-c c'\right)
-c^2 \dot{\mathcal{K}}' 
-\frac{1}{2} \ell (\ell+1) \dot{\mathcal{H}}_1 =0 \, ,
\label{eqmcH1}
\end{equation}
which can be solved algebraically for $H_1$. Plugging the field redefinition \eqref{psi} together with the solutions for $\mathcal{H}_1$ and $H_1$ into the effective theory \eqref{bHorn-2} and eliminating higher derivative terms that may be present through suitable integrations by parts, one finds that the quadratic action for the physical even parity modes takes the form
\begin{equation} \label{even quadratic action}
	S^{(2)}_{\text{even}, m=0} =  \frac{\Mpl^2}{2} \sum_{\ell\geq2} \int \D t \D r \left( \mathcal{A}^{ij} \dot \chi_i \dot \chi_j - \mathcal{B}^{ij} \chi_i' \chi_j' - \mathcal{C}^{ij} \chi_i \chi_j' - \mathcal{D}^{ij} \chi_i \chi_j \right),
\end{equation}
where $\chi\equiv (\psi, \mathcal{K})$ and
where we have restricted our attention to the modes with $m=0$, in such a way that the fields' perturbations are real. This is possible in general because the spherical symmetry of the background guarantees that modes with different values of $m$ will satisfy the same equations of motion.
The explicit  expressions for $\mathcal{A}$ and $\mathcal{B}$ are:
\begin{subequations}\label{Aij}
\begin{align}
	\mathcal{A}_{\psi\psi}  & = 
	{\scriptsize \text{ $
-8a \bigg[  \Mpl^2 c'^2 \left(\left(-\ell^2-\ell+2\right) \Mpl^2+4 \ell(\ell+1) M_{13}^2\right)+\Mpl^2 c \left(4 M_3^3 c'+\ell(\ell+1) \left(c'' \left(\Mpl^2-4 M_{13}^2\right)+4 c' (M_{13}^2)'\right)\right)
	$}} \nonumber
\\
& {\scriptsize \text{ $
+c^2 \left(4 \ell(\ell+1) M_3^3 (M_{13}^2)'+\ell (\ell+1) \left(\Mpl^2-4 M_{13}^2\right) (M_3^3)'+2 (M_3^3)^2\right)
- \frac{c a'}{a}\ell(\ell+1)  \left(\Mpl^2-4 M_{13}^2\right) \left(\Mpl^2 c'+c M_3^3\right) \bigg]
$}} \nonumber
\\
& {\scriptsize \text{ $
\times \bigg[ \ell(\ell+1) \left(2 c a' \left(\Mpl^2 c'+c M_3^3\right)+a \left(\Mpl^2 \left(-2 c'^2+\ell^2+\ell\right)-2 c M_3^3 c'-4 \ell (\ell+1) M_{13}^2\right)\right)^2 \bigg]^{-1}
$} } \, ,
	\\
\mathcal{A}_{\psi\mathcal{K}}  & = \mathcal{A}_{\mathcal{K}\psi} =
	{\scriptsize \text{ $
-\frac{2c\left(\ell^2+\ell-2\right) \left(\Mpl^2 c'+c M_3^3\right)}{\ell(\ell+1) \left(2 c a' \left(\Mpl^2 c'+c M_3^3\right)+a \left(\Mpl^2 \left(-2 c'^2+\ell^2+\ell\right)-2 c M_3^3 c'-4 \ell (\ell+1) M_{13}^2\right)\right)}
$} } \, ,
	\\
\mathcal{A}_{\mathcal{K}\mathcal{K}}  & =
	{\scriptsize \text{ $
c^2\frac{\ell^2+\ell-2}{2a \ell(\ell+1)}
$} } \, ,
\end{align} 
\end{subequations}
\begin{subequations}\label{Bij}
\begin{align}
	\mathcal{B}_{\psi\psi}  & = 
	{\scriptsize \text{ $
-8a^3 \bigg[ \frac{c a' c'}{a} \ell (\ell+1) \Mpl^4 + \left(\ell^2+\ell-2\right) \Mpl^4 c'^2-c \left(\ell (\ell+1) \Mpl^4 c''
-2 \left(\ell^2+\ell-2\right) \Mpl^2 M_3^3 c'\right)
	$}} \nonumber
\\
& {\scriptsize \text{ $
-2 c^2 \left(\left(-2 \ell^2-2 \ell+1\right) (M_3^3)^2+2 \ell (\ell+1) \Mpl^2 M_2^4\right) \bigg]
$}} \nonumber
\\
& {\scriptsize \text{ $
\times \bigg[ \ell(\ell+1) \left(2 c a' \left(\Mpl^2 c'+c M_3^3\right)+a \left(\Mpl^2 \left(-2 c'^2+\ell^2+\ell\right)-2 c M_3^3 c'-4 \ell (\ell+1) M_{13}^2\right)\right)^2 \bigg]^{-1}
$} } \, ,
	\\
\mathcal{B}_{\psi\mathcal{K}}  & = \mathcal{B}_{\mathcal{K}\psi} =  a^2 \mathcal{A}_{\psi\mathcal{K}} \, ,
	\\
\mathcal{B}_{\mathcal{K}\mathcal{K}}  & =  a^2  \mathcal{A}_{\mathcal{K}\mathcal{K}} \, .
\end{align} 
\end{subequations}
Diagonalizing the matrix $c_r^2 = \frac{1}{a^2}\mathcal{A}^{-1}\mathcal{B}$ with $\mathcal{A}$ and $\mathcal{B}$ given in \eqref{Aij} and \eqref{Bij}, one can easily compute the speeds of propagation along the radial direction:
\begin{equation}
c_{r,1}^2  = 
a\frac{\Mpl^2 \left(\frac{a' c'}{ac} -  \frac{c''}{c}\right) - 4  M_2^4 + 3 \frac{(M_3^3)^2}{\Mpl^2}}{ 
\left(M_3^3+\Mpl^2\frac{c'}{c} \right)^2
\left[ \frac{\partial}{\partial r} \left( \frac{ac\left(M_\text{Pl}^2-4M_{13}^2 \right)}{M_3^3c +M_\text{Pl}^2c'} \right)	-  a   \right] 
}
 \, ,
\qquad  c_{r,2}^2  = 1 \, .
  \label{cs1}
\end{equation}
We do not report here the expressions for $\mathcal{C}$ and $\mathcal{D}$, since they are quite involved and irrelevant for the purposes of discussing ghost/gradient instabilities.
Instead, in order to discuss the absence of ghost and gradient instabilities along the radial direction, we focus  on the kinetic matrices $\mathcal{A}$ and $\mathcal{B}$, and demand that they are positive definite.
Looking more closely at the expressions \eqref{Aij} and \eqref{Bij},
one infers immediately that, since $\ell\geq2$, $\mathcal{A}_{\mathcal{K}\mathcal{K}}>0$ and $\mathcal{B}_{\mathcal{K}\mathcal{K}}>0$. Thus, given that $\mathcal{A}$ and $\mathcal{B}$ are symmetric, the positive definiteness simply follows from the positivity of their determinants, which read\footnote{As a check, one can notice that the ratio $\frac{\det(\mathcal{B})}{a^2\det(\mathcal{A})}$ coincides with the product of the two eigenvalues in \eqref{cs1}.} 
\begin{multline}
\det(\mathcal{A}) =
\\
= {\scriptsize \text{ $
\frac{\frac{4c^2}{a} \frac{\left(\ell^2+\ell-2\right)}{\ell(\ell+1)} \left(\Mpl^2 c'+c M_3^3\right)^2 }{ \left[2 c a' \left(\Mpl^2 c'+c M_3^3\right)+a \left(\Mpl^2 \left(\ell^2+\ell-2 c'^2\right)-2 c M_3^3 c'-4 \ell(\ell+1) M_{13}^2\right)\right]^2}
\left[ \frac{\partial}{\partial r} \left( \frac{ac\left(M_\text{Pl}^2-4M_{13}^2 \right)}{M_3^3c +M_\text{Pl}^2c'} \right)	-  a   \right] 
$} } >0  \, ,
\label{detA}
\end{multline} 
\begin{equation}
\det(\mathcal{B})
= {\scriptsize \text{ $
\frac{4a^4c^4 \frac{\left(\ell^2+\ell-2\right)}{\ell(\ell+1)}\left[ \Mpl^4\left( \frac{a' c'}{ac} -  \frac{c''}{c} \right) - 4 \Mpl^2 M_2^4 + 3 (M_3^3)^2\right]  }{ \left[2 c a' \left(\Mpl^2 c'+c M_3^3\right)+a \left(\Mpl^2 \left(\ell^2+\ell-2 c'^2\right)-2 c M_3^3 c'-4 \ell(\ell+1) M_{13}^2\right)\right]^2}
$} } >0 \, .
\label{detB}
\end{equation} 
As already known in the context of Horndeski theories, there is no particular obstruction in fulfilling the second condition \eqref{detB}: one can choose for instance $M_2^4(r)$ in such a way that Eq. \eqref{detB} is satisfied for all $r$, preventing the occurrence of gradient instabilities along the radial direction. 

Let us now focus on the positivity condition \eqref{detA} for the kinetic matrix.
In parallel with the considerations outlined in \cite{Creminelli:2016zwa,Cai:2016thi} in the context of geodesically complete cosmologies, it is convenient to introduce the variable
\begin{equation}
Y \equiv \frac{ac\left(M_\text{Pl}^2-4M_{13}^2 \right)}{M_3^3c+M_\text{Pl}^2c'}  \, ,
\label{Y}
\end{equation}
in such a way that Eq. \eqref{detA} now reads
\begin{equation}
Y'(r) > a(r) > 0 \, .
\label{monotonic}
\end{equation}
From Eq. \eqref{monotonic} one infers, first of all, that $Y$ has to be a monotonically growing function of $r$. Furthermore, integrating in $r$, 
\begin{equation}
Y(r_+) - Y(r_-) > \int_{r_-}^{r_+} \D r \, a(r) \, .
\label{condstab}
\end{equation}
The asymptotic conditions \eqref{backm2} make the integral on the r.h.s. of \eqref{condstab} divergent on both ends when $r_\pm \to \pm \infty$. Thus, the inequality \eqref{condstab} forces $Y(r_{\pm}\rightarrow\pm\infty)\rightarrow\pm\infty$. Together with the monotonicity \eqref{monotonic}, this implies that, barring discontinuities, the function $Y$ must vanish at a single point $r_0$, corresponding to
\begin{equation}
M_{13}^2(r_0) = \frac{M_\text{Pl}^2}{4} \, .
\label{condM13}
\end{equation}
It should be clear now that, without the operator $\delta g^{rr } \delta\hat{R} $ (\textit{i.e.} if $M_{13}^2=0$), the function $Y(r)$ would be nonzero for all $r$ (see Eq. \eqref{Y}), invalidating the inequality \eqref{condstab} at least in a certain interval and resulting in an unavoidable ghost-like instability in agreement with \cite{Rubakov:2015gza,Rubakov:2016zah}.\footnote{It is worth noticing that the inclusion of additional matter fields would not change this result. See the discussion in \cite{Creminelli:2016zwa} about this point.} Thus, in order to find stable wormhole solutions one needs to go beyond the standard Horndeski class and consider \textit{e.g.} theories of the type \eqref{bHorn-2}.

\section{Conclusions and outlook}
\label{concls}

Taking advantage of the effective description introduced in \cite{Franciolini:2018uyq} for perturbations around spherically symmetric backgrounds, we have reconsidered the no-go theorems \cite{Rubakov:2015gza,Rubakov:2016zah,Evseev:2017jek} that prevent the existence of stable wormhole solutions in the case of Horndeski theories, capturing in general terms the origin of the pathology. Moreover, we have identified at least one operator in the unitary gauge that is necessary to overcome the issue, emphasizing that it belongs to the class of the so-called beyond-Horndeski theories \cite{Gleyzes:2014dya,Zumalacarregui:2013pma}. 

The analysis has worked in parallel with the one presented in \cite{Creminelli:2016zwa,Cai:2016thi} for geodesically complete FLRW backgrounds. In both cases a linearly stable NEC-violating geometry can be obtained only in a theory with higher derivative equations of motion which, in spite of that, does not propagate any additional pathological degree of freedom. The crucial question is whether such a peculiar dynamics, that seems well-defined and healthy from an EFT point of view, can be UV completed in a Lorentz invariant theory with an S-matrix that satisfies the standard analyticity conditions.   

More prosaically, it would be interesting to find an explicit example of covariant Lagrangian that allows for a wormhole solution of the form we are considering and, at the same time, generates the quadratic operator in the EFT for perturbations that is needed for stability. This will be presented in a forthcoming paper.


\section*{Acknowledgements}

L.S. is supported by Simons Foundation Award Number 555117. E.T. is partially supported by the MIUR PRIN project 2015P5SBHT. L.H. is supported in part by NASA grant NXX16AB27G and DOE grant DE-SC011941.


\appendix

\section{Quartic beyond-Horndeski in unitary gauge}
\label{Sec:conf-trans}

In this section we expand in perturbations the unitary gauge quartic beyond-Horndeski Lagrangian \cite{Gleyzes:2014dya,Zumalacarregui:2013pma} and make use of the conformal and disformal transformations \cite{Bekenstein:1992pj} briefly reviewed in App. \ref{app:trans} below to simplify the final expression. In particular, we shall see that a disformal transformation is enough to remove both the combination $\delta K^2-(\delta K_{\mu\nu})^2$ and the operator $\delta N \delta K^{\mu\nu}$, while a conformal transformation can be used to recover the Einstein frame.
The starting point is an action in the form
\begin{equation}
S =  \int\D^4x  \sqrt{-g}  \left[
	A(r,N)\hat{R} 
	+ B(r,N)\left( K^2-K_{\mu\nu}K^{\mu\nu} \right) 
\right]  \, ,
\label{0heb}
\end{equation}
where $A$ and $B$ are arbitrary functions of the radial coordinate $r$ and the lapse $N=\frac{1}{\sqrt{g^{rr}}}$.
Let us consider a field redefinition of the type \eqref{0ct} in App. \ref{app:trans} where $\Omega$ is taken to be a function of $r$ only. The transformation laws \eqref{tK}-\eqref{0cf1} take on the form  
\begin{equation}
{K^a}_{b} \mapsto {\tilde{K}^a}_{\, \, \, b}
= \frac{N}{\tilde{N}}\left( {K^a}_{b} +  \frac{\Omega'}{N \Omega}\delta^a_b    \right)\, ,
\qquad
\hat{R} \mapsto \tilde{\hat{R}} = \Omega^{-2}\hat{R} \, ,
\end{equation}
where the latin indices $a, b, ...$ run over the $(t, \theta, \phi)$-components.
Then the full action \eqref{0heb} transforms as
\begin{multline}
S =  \int\D^4x  \sqrt{-g} \,  \Omega^2 \sqrt{1+\frac{\Gamma}{\Omega^2}} \bigg[
	\tilde{A}(r,N) \hat{R}
\\
	+ \frac{1}{1+\frac{\Gamma}{\Omega^2}} \tilde{B}(r,N) \left(K^2-K_{\mu\nu}K^{\mu\nu} + \frac{4\Omega'}{N\Omega}K
	+ \frac{6\Omega'^2}{N^2\Omega^2} \right) 
\bigg]  \, .
\label{20heb-2}
\end{multline}
Expanding now in perturbations up to quadratic order, we find
\begin{multline}
S^{(2)} =  \int\D^4x  \sqrt{-g} \, \Omega^2 \Bigg[
	 \sqrt{1+\frac{\Gamma}{\Omega^2}}  \, \tilde{A}(r,\bar{N}) \hat{R}
	+ \frac{\tilde{B}(r,\bar{N})}{\sqrt{1+\frac{\Gamma}{\Omega^2}} } \left(K^2-K_{\mu\nu}K^{\mu\nu} \right)
\\
	+  \left( \sqrt{1+\frac{\Gamma}{\Omega^2}}  \, \tilde A\right)_{,N}  \delta N \delta \hat{R}
	-2\left(  \frac{ \tilde{B}}{\sqrt{1+\frac{\Gamma}{\Omega^2}} } \right)_{,N}  \bar K_{\mu\nu} \delta N \delta K^{\mu\nu}  + \ldots
\Bigg]  \, ,
\label{20heb-3}
\end{multline}
where in the dots we are dropping terms that contribute to the operators $\delta N$, $\delta N^2$, $\delta N \delta K$ and the tadpole $\Lambda(r)$, which are not interesting for the purpose of this section. Notice that in \eqref{20heb-3} every coefficient should be read of as being computed on the background.

Now it is clear from \eqref{20heb-3} that one is free to choose $\Omega(r)$, $\Gamma(r,\bar{N})$ and $\Gamma_{,N}(r,\bar{N})$ in such a way that
\begin{equation}
\Omega^2(r)\tilde{A}(r,\bar{N})\sqrt{\frac{\tilde B(r,\bar{N})}{\tilde A(r,\bar{N})}} = \frac{\Mpl^2}{2} \, ,
\end{equation}
\begin{equation}
\Gamma(r,\bar N) = \Omega^2(r) \left[ \frac{\tilde B(r,\bar{N})}{\tilde A(r,\bar{N})} -1 \right] \, ,
\quad
\Gamma_{,N}(r,\bar N) = \frac{2\Omega^2(r)}{\tilde{A}(r,\bar N)} \tilde{B}_{,N} \, ,
\end{equation}
so that the coefficients of both $\delta K^2-(\delta K_{\mu\nu})^2$ and $\delta N \delta K^{\mu\nu}$ are set to zero, while restoring the Einstein frame. Therefore, the final result has the form \eqref{bHorn-2}.

We conclude this section commenting on the fact that, by contrast, one can not remove in full generality the operator $\delta N \delta \hat{R}$.\footnote{That this has to be the case can be also  inferred from the analysis of Sec. \ref{sec:eftw}, which revealed that the stability of the wormhole solution, which is a frame-independent property, crucially depends on the presence of the operator $\delta N \delta\hat{R}$.}
To illustrate the argument we start from the action in the form \eqref{bHorn-2} and consider a pure disformal transformation ($\Omega\equiv 1$) such that $\Gamma(r,\bar N)\equiv 0$. This time we shall try to fix  $\Gamma_{,N}(r,\bar{N})$  in such a way to set to zero the coefficient of $\delta N \delta \hat{R}$ instead of $\delta N \delta K^{\mu\nu}$.
After a straightforward calculation, one finds that
\begin{equation}
M_{13}^2 (r) \mapsto \tilde{M}_{13}^2(r) = \frac{M_{13}^2(r) + \frac{\Mpl^2}{8}\bar N^3\Gamma_{,N}(r,\bar N)}{1+ \frac{1}{2}\bar N^3\Gamma_{,N}(r,\bar N)} \, .
\label{disfMb60}
\end{equation}
Setting $\tilde{M}_{13}^2\equiv 0$ in the new frame is equivalent to choosing
\begin{equation}
\frac{1}{2}\bar N^3\Gamma_{,N}(r,\bar N) = - \frac{M_{13}^2(r)}{\Mpl^2/4} \, .
\label{disfMb6}
\end{equation}
Now, according to the discussion of Sec. \ref{sec:eftw}, a necessary condition to avoid ghost instabilities in the  class of theories \eqref{bHorn-2} is that the coefficient of $\delta g^{rr} \delta \hat{R}$ equals $\Mpl^2/4$ at a certain distance---see Eq. \eqref{condM13}. In other words,  there must exist a point $r_0$ where the r.h.s. of Eq. \eqref{disfMb6} is $-1$, which in turn makes the transformation itself ill defined, as clear from Eq. \eqref{disfMb60}.
As a result, in agreement with the findings of Sec. \ref{sec:eftw}, one can not remove $\delta g^{rr} \delta \hat{R}$ everywhere along the trajectory of the wormhole background solution. Notice that there exists an analogous argument in the context of geodesically complete FLRW spacetimes \cite{Creminelli:2016zwa}.

\section{Conformal and disformal transformations}
\label{app:trans}

Let us consider now the following field redefinition \cite{Bekenstein:1992pj}:
\begin{equation}
g_{\mu\nu} \mapsto \tilde{g}_{\mu\nu} = \Omega^2(r,N)g_{\mu\nu} + \Gamma(r,N)n_\mu n_\nu \, ,
\label{0ct}
\end{equation}
where $\Omega$ and $\Gamma$ are functions of the coordinate $r$ and the lapse $N=\frac{1}{\sqrt{g^{rr}}}$. Denoting with latin indices $a, b, ...$ the $(t, \theta, \phi)$-components, the relevant geometric quantities transform accordingly as follows\footnote{For FLRW backgrounds see, for instance, Sec. 3.4 of \cite{Gleyzes:2014rba}.}:
\begin{equation}
N_a \mapsto \tilde{N}_a = \Omega^2  N_a \, ,
\qquad
N^a \mapsto \tilde{N}^a = N^a \, ,
\qquad
N^2 \mapsto \tilde{N}^2 = (\Omega^2 +\Gamma) N^2 \, ,
\end{equation}
\begin{equation}
{K^a}_{b} \mapsto {\tilde{K}^a}_{\, \, \, b} = \frac{N}{\tilde{N}}\left( {K^a}_{b} + \delta^a_b N g^{r\mu}\partial_\mu \ln \Omega \right)
= \frac{N}{\tilde{N}}\left[ {K^a}_{b} +  \frac{\delta^a_b}{N \Omega} \left( \Omega_{,r} + \Omega_{,N}\left(N' - N^a\partial_aN \right)   \right)  \right] \, ,
\label{tK}
\end{equation}
\begin{equation}
\hat{R} \mapsto \tilde{\hat{R}} = \Omega^{-2}\left(
\hat{R} -4D_a D^a\ln \Omega
-2\partial_a\ln\Omega \, \partial^a\ln\Omega
\right)  \, ,
\label{0cf1}
\end{equation}
\begin{equation}
\sqrt{-g} \mapsto \sqrt{-\tilde{g}} = \sqrt{-g} \,  \Omega^3 \sqrt{\Omega^2 +\Gamma} \, ,
\end{equation}
while any generic scalar function transforms as $f(r,N)\mapsto \tilde{f}(r,N)\equiv f ( r, \tilde{N}(r,N))$. 
A rapid inspection of the previous expressions reveals that the transformation \eqref{0ct} is ill defined on the domain where $\Gamma=-\Omega^2$, which corresponds to a singularity in the metric tensor. Thus, we will assume in the following that $\Gamma\neq-\Omega^2$ everywhere.

\section{More on stability conditions}
\label{App:l0l1}

In this appendix we complete the study of the wormhole's stability by considering perturbations with $\ell=1$ and $\ell=0$, which have been disregarded in the main text. As we shall see, one recovers the same  condition \eqref{condstab} for the absence of ghost instabilities.

\paragraph{Stability conditions for modes with $\ell=1$.} Plugging the explicit expressions for the spherical harmonics $Y_{1m}$ (with $m=0, \pm1$) into Eq. \eqref{even metric perturbations} one finds that the dipole perturbations $\mathcal{K}$ and $G$ appear only through the combination $\mathcal{K}-G$. This means that one is free to set either $\mathcal{K}$ or $G$ to zero from the outset. On top of that, one can use the residual gauge freedom to get rid of other two components in $\delta g_{\mu\nu}^\text{even}$. Ultimately, we can work in the following gauge
\begin{equation}
\mathcal{K}=G=\mathcal{H}_0 =0 \, .
\end{equation}
The computation of the quadratic action for the physical mode closely follows the steps outlined in Sec. \ref{sec:l2} for the case $\ell\geq2$.

The constraint equation obtained from the variation with respect to the Lagrange multiplier $H_0$ can be read off from Eq. \eqref{LM1} upon setting $\ell=1$ and $\mathcal{K}=0$. 
Then, the analogous field redefinition
\begin{equation}
H_2 \equiv \left( \frac{  c c' }{2} + \frac{  c^2 M_3^3}{2 \Mpl^2} \right)^{-1}
\left(
\psi +  \mathcal{H}_1 
\right) \, ,
\end{equation}
can be used to get rid of $\mathcal{H}'_1$ and make the equation algebraic in $\mathcal{H}_1$.  Similarly, the equation of motion for the non-dynamical component $H_1$ is given in Eq. \eqref{eqmcH1} with $\ell=1$ and $\mathcal{K}=0$. This provides an expression for  $H_1$ in terms of $\dot{H}_2$ and $\dot{\mathcal{H}}_1$. Plugging the expressions for $\mathcal{H}_1$ and $H_1$ into the effective theory \eqref{bHorn-2} yields the following quadratic action for the propagating scalar mode:
\begin{equation} \label{even quadratic actionl1}
	S^{(2)}_{\text{even}, \ell=1,m=0} =  \frac{\Mpl^2}{2} \int \D t \D r \left( A_1 \dot{\psi}^2  - B_1 \psi'^2 - C_1 \psi^2 \right) \, ,
\end{equation}
where 
\begin{equation}
A_1
= {\scriptsize \text{ $
\frac{2 \left(\Mpl^2 c'+c M_3^3\right)^2 }{ \left[c a' \left(\Mpl^2 c'+c M_3^3\right)+a \left(\Mpl^2 \left(1- c'^2\right)- cc' M_3^3 -4 M_{13}^2\right)\right]^2}
\left[ \frac{\partial}{\partial r} \left( \frac{ac\left(M_\text{Pl}^2-4M_{13}^2 \right)}{M_3^3c +M_\text{Pl}^2c'} \right)	-  a   \right] 
$}}  \, ,
\label{Al1}
\end{equation} 
while $B_1= a^2 A_1 c_{r,1}^2$, where $c_{r,1}^2$ is given in Eq. \eqref{cs1}.
Thus, it is clear from the previous expressions that the analysis of stability goes unaltered with respect to the case $\ell\geq2$.

\paragraph{Stability conditions for modes with $\ell=0$.} Since $Y_{00}$ is a constant, the monopole perturbations $\mathcal{H}_0$, $\mathcal{H}_1$ and $G$ automatically disappear  from the metric tensor  \eqref{even metric perturbations}. Furthermore, in the unitary gauge $\delta\Phi\equiv0$ one is still free to fix the residual time diffeomorphisms in such a way that e.g. $H_0=0$. Then, straightforward integrations by parts lead to the cancellation of the quadratic term for $H_1$, which becomes therefore a Lagrange multiplier enforcing a differential equation for $\psi$ and $\mathcal{K}$ whose solution reads
\begin{equation}
\mathcal{K} = \frac{2 }{c \left(c a'-a c'\right)} \psi \, ,
\label{solKl0}
\end{equation}
where we have set to zero an inconsequential integration constant.\footnote{Had we chosen to set to zero $H_1$ instead of $H_0$, we would have found a non-independent constraint equation with the same solution \eqref{solKl0}.}
Plugging back into the action \eqref{bHorn-2}, one finds the quadratic Lagrangian for the scalar mode to be
\begin{equation} \label{even quadratic actionl0}
	S^{(2)}_{\text{even}, \ell=0,m=0} =  \frac{\Mpl^2}{2} \int \D t \D r \left( A_0 \dot{\psi}^2  - B_0 \psi'^2 - C_0 \psi^2 \right) \, ,
\end{equation}
where 
\begin{equation}
A_0
= 
\frac{2 }{\left(c a'-a c'\right)^2}
\left[ \frac{\partial}{\partial r} \left( \frac{ac\left(M_\text{Pl}^2-4M_{13}^2 \right)}{M_3^3c +M_\text{Pl}^2c'} \right)	-  a   \right] 
  \, ,
\label{Al0}
\end{equation} 
while $B_0= a^2 A_0 c_{r,1}^2$.


\bibliographystyle{utphys}
\bibliography{EFTofSSB}

\end{document}